\documentclass[conference]{IEEEtran}
\IEEEoverridecommandlockouts

\usepackage{cite}
\usepackage{amsmath,amssymb,amsfonts}
\usepackage{algorithm}
\usepackage{algorithmic}
\usepackage{graphicx}
\usepackage{textcomp}
\usepackage{xcolor}
\usepackage{booktabs,array}
\usepackage{placeins}
\def\BibTeX{{\rm B\kern-.05em{\sc i\kern-.025em b}\kern-.08em
    T\kern-.1667em\lower.7ex\hbox{E}\kern-.125emX}}

\begin{document}

\title{Robust Constraint-Aware Bayesian Tuning of BBRv2 for QUIC under Tactile Internet Constraints}

\author{
\IEEEauthorblockN{Muhammad Hanif Lashari$^1$,  Shakil Ahmed$^2$, Wafa Batayneh$^1$, and Ashfaq Khokhar$^1$}
\IEEEauthorblockA{
$^1$Department of Electrical and Computer Engineering, Iowa State University, Iowa, USA \\
$^2$Department of Computer Science, College of Computing, Grand Valley State University, Michigan, USA \\
Email: mhanif@iastate.edu, ahmeshak@gvsu.edu, \{batayneh, ashfaq\}@iastate.edu\\
(This work will appear in Globecom 2026 IoTSN)
}
}

\maketitle

\begin{abstract}
Tactile Internet applications place strict requirements on latency, jitter, loss, and responsiveness, which makes transport configuration a critical design factor. Although BBRv2 offers a model-based congestion control framework with strong throughput potential, its default behavior may not be well aligned with delay-sensitive interactive scenarios. This paper presents a robust and constraint-aware tuning framework for BBRv2 in QUIC, where parameter selection is formulated as an expensive black-box optimization problem over multiple emulated network conditions. The tuning process uses Bayesian optimization with the Tree Structured Parzen Estimator to efficiently explore a bounded parameter space under noisy experimental measurements. The objective is designed to preserve throughput while enforcing limits on tail latency and loss, while delay instability is evaluated separately through the jitter metric. Experimental results across low, medium, and high impairment scenarios show that the tuned configuration improves tail latency, jitter behavior, and loss performance while maintaining competitive goodput relative to standard QUIC congestion control baselines. These results support robust black-box tuning as a practical method for adapting QUIC transport behavior to tactile Internet style requirements.
\end{abstract}


\section{Introduction}
\label{sec:introduction}

Tactile Internet (TI) applications aim to support real-time haptic and control loops with strict quality of service requirements on latency, jitter, and reliability~\cite{2chaudhari2025enabling}.
Standardization work on TI quality of service explicitly emphasizes assurance requirements and application-driven constraints rather than best effort delivery~\cite{ITUY3149}.
Remote robotic assisted procedures are a representative high-stakes use case where network latency, jitter, and packet loss can impact safe operation and perceived responsiveness~\cite{wang2025expert}.

Ultra reliable low latency communication has been introduced to address mission critical applications where sub millisecond latency targets and very high reliability are discussed in the 5G and 6G context~\cite{maghsoudnia2024ultra,ITUY3149}.
Practical analyses argue that achieving these targets requires system level design choices because every source of delay becomes a bottleneck at sub millisecond regimes~\cite{maghsoudnia2024ultra}.
These requirements motivate transport configurations that reduce queue buildup and avoid delay spikes rather than optimizing average throughput alone~\cite{2chaudhari2025enabling,ITUY3149,lashari2025predictive}.

QUIC is a standardized secure transport protocol that runs over UDP and integrates cryptographic handshake and transport negotiation~\cite{RFC9000}.
Because QUIC is commonly implemented in user-space, it supports rapid iteration and deployment of congestion control logic compared with kernel-based transports~\cite{RFC9000,StarQUIC2024}.
Although user-space processing may introduce scheduling and
memory-access overhead, QUIC enables rapid modification and
deployment of congestion-control logic without kernel changes.
Accordingly, this work evaluates transport-level congestion behavior rather than complete end-to-end tactile-loop latency.

Recent experimental work continues to study how QUIC performance varies with congestion control choices and queue management policies under emulated conditions~\cite{Bujari2024QUICCCA}.

Congestion control behavior in QUIC depends on the selected algorithm and its configuration~\cite{Bujari2024QUICCCA,StarQUIC2024}.
Loss-based approaches such as Reno and CUBIC remain widely used baselines, while model-based approaches such as BBR and BBRv2 attempt to regulate sending using estimates of bottleneck bandwidth and propagation delay~\cite{Abrol2026Survey,Cardwell2024BBRDraft}.
Surveys and recent evaluations highlight that parameter choices and implementation details can significantly alter throughput, loss, and queue behavior for BBR family algorithms~\cite{Abrol2026Survey,Sarpkaya2025IFIP}.

This paper studies robust tuning of BBRv2 for QUIC under TI constraints.
The core challenge is that the mapping from parameters to measured performance is not available in closed form and can only be observed by running experiments~\cite{Diessner2025Thesis,Chen2024PEI}.
This yields an expensive, noisy black-box optimization problem where sample efficiency is essential because each evaluation has nontrivial cost~\cite{Diessner2025Thesis,GonzalezDuque2024NeurIPS}.
\subsection{Contributions}
\label{subsec:contributions}
The first contribution is a robust, constraint-aware problem formulation for BBRv2 tuning in QUIC across multiple network scenarios rather than one fixed profile~\cite{StarQUIC2024}.
The second contribution is a practical black-box tuning framework based on Bayesian optimization using the Tree Structured Parzen Estimator for bounded search spaces with noisy outcomes~\cite{Bergstra2011TPE,Watanabe2023cTPE,Diessner2025Thesis}.
The third contribution is an evaluation plan that compares the tuned configuration against established baselines using throughput, tail latency, loss, and delay stability metrics aligned with TI assurance requirements~\cite{ITUY3149,Bujari2024QUICCCA}.

\section{Problem Motivation}
\label{sec:motivation}

\subsection{Need for Robust Parameter Tuning}
\label{subsec:motivation_tuning}

BBRv2 exposes multiple gain parameters and thresholds that control pacing, probing, and congestion window behavior~\cite{Abrol2026Survey,Cardwell2024BBRDraft}.
Recent studies emphasize that small heuristic and implementation changes can significantly alter BBR behavior and coexistence outcomes, which makes default settings sensitive to deployment context~\cite{Sarpkaya2025IFIP,Abrol2026Survey}.
Empirical evaluations under diverse environments continue to show that congestion control behavior and recommended configurations depend on queue policy, delay, loss, and path dynamics~\cite{Bujari2024QUICCCA,StarQUIC2024}.

The TI requirements prioritize tail latency and stability in addition to throughput. Remote procedure guidelines explicitly list low latency, minimal jitter, minimal packet loss, and guaranteed bandwidth as key network characteristics for safe operation.
These requirements motivate tuning that targets delay spikes and instability, since a configuration can have high throughput while still producing unacceptable delay excursions~\cite{ITUY3149}.

Robustness across scenario variation is necessary because tactile deployments can experience different delay, bandwidth, queue, and loss settings across locations and operating conditions.
Tuning on a single fixed profile risks overfitting and can fail when conditions shift even moderately~\cite{StarQUIC2024,Bujari2024QUICCCA}.
A scenario set based robust objective better matches the requirement to deliver stable behavior across a range of conditions~\cite{ITUY3149}.

\subsection{Black-Box Nature of the Tuning Problem}
\label{subsec:motivation_blackbox}

The relationship between congestion control parameters and measured performance is difficult to model analytically because the outcome depends on packet level feedback, timers, state transitions, and implementation details~\cite{Abrol2026Survey,Cardwell2024BBRDraft}.
In QUIC, performance metrics such as goodput, tail latency, and loss are observed only after running an experiment under a specific network profile~\cite{Bujari2024QUICCCA,StarQUIC2024}.
This makes the objective expensive and opaque, which is a defining property of an expensive black-box optimization problem~\cite{Diessner2025Thesis,Chen2024PEI}.

Noisy measurements are expected because repeated network experiments can vary due to timing variability, background effects, and protocol dynamics~\cite{Bujari2024QUICCCA,Diessner2025Thesis}.
The black-box optimizers are designed for settings where gradients are unavailable or unreliable and where only a limited number of evaluations can be afforded~\cite{Diessner2025Thesis,Chen2024PEI}.
Bayesian optimization is commonly used for expensive black-box objectives because it aims to reduce evaluation count by using past trials to guide new proposals~\cite{Diessner2025Thesis,GonzalezDuque2024NeurIPS}.

The Tree Structured Parzen Estimator is a practical Bayesian optimization approach that builds models of promising and non-promising regions and proposes new candidates accordingly~\cite{Bergstra2011TPE,Watanabe2023cTPE}.
Recent work extends TPE to inequality constrained settings, which aligns with constraint-aware tuning where feasibility matters~\cite{Watanabe2023cTPE}.
This motivates using TPE as the core search engine for bounded parameter spaces with noisy measurements and hard limits~\cite{Bergstra2011TPE,Watanabe2023cTPE}.

\subsection{Need for Constraint-Aware Optimization}
\label{subsec:motivation_constraints}

Tactile Internet assurance requirements emphasize strict bounds on latency, jitter, and reliability rather than average performance alone.
Remote robotic assisted procedure guidelines explicitly highlight that low latency, minimal jitter, and minimal to no packet loss are essential to prevent delays that could compromise safe operation.
These sources motivate treating tail latency and loss as limits that must be satisfied, not as soft preferences.

Constraint-based formulations also provide a clear interpretation for system design because they separate mandatory requirements from optimization preferences~\cite{Diessner2025Thesis}.
Penalty-based black-box objectives are commonly used to convert constrained problems into a single score that can be minimized by general optimizers~\cite{Chen2024PEI,Watanabe2023cTPE}.
This approach fits the tuning setting where constraint violations must be strongly discouraged while still allowing throughput maximization within feasible regions~\cite{ITUY3149,Watanabe2023cTPE}.

\section{Related Work}
\label{sec:related_work}
\subsection{QUIC Congestion Control Studies}
\label{subsec:rw_quic}

QUIC is standardized as a secure general purpose transport protocol over UDP with integrated cryptographic handshake~\cite{RFC9000}.
Recent studies continue to evaluate QUIC performance under different congestion control algorithms and queue management policies using emulation~\cite{Bujari2024QUICCCA,StarQUIC2024}.
Such evaluations highlight that transport outcomes depend on both the congestion control algorithm and the queue behavior at the bottleneck~\cite{Bujari2024QUICCCA}.

\subsection{BBR Variants and Tuning Sensitivity}
\label{subsec:rw_bbr}

BBR and its later versions are model-based congestion control algorithms that aim to regulate sending using estimates of bottleneck bandwidth and propagation delay rather than using loss as the primary congestion signal~\cite{Cardwell2024BBRDraft,Abrol2026Survey}.
The IETF congestion control work documents how BBR modulates pacing gain and congestion window gain across its state machine~\cite{Cardwell2024BBRDraft}.
A recent comprehensive survey summarizes the evolution of BBRv1, BBRv2, and BBRv3 and discusses performance and fairness concerns across environments~\cite{Abrol2026Survey}.

Experimental and modeling work shows that BBR coexistence with loss-based flows can depend on version details and buffer settings, and that small heuristic changes can shift behavior~\cite{Sarpkaya2025IFIP,Abrol2026Survey}.
This sensitivity motivates tuning frameworks that are explicit about scenario coverage and constraints rather than relying on a single default configuration~\cite{Abrol2026Survey,StarQUIC2024}.
Recent work on QUIC over LEO networks explicitly discusses hyperparameter tuning for BBRv2 in response to high loss and latency variation~\cite{StarQUIC2024}.

\subsection{Black-Box Optimization for Systems Tuning}
\label{subsec:rw_blackbox}

Optimization of expensive black-box objectives is widely studied in engineering contexts where evaluations require high fidelity simulation or physical experiments~\cite{Chen2024PEI,Diessner2025Thesis}.
Bayesian optimization is a common strategy in this setting because it uses surrogate models and acquisition rules to select promising candidates sequentially under limited evaluation budgets~\cite{Diessner2025Thesis,GonzalezDuque2024NeurIPS}.
Recent benchmark and survey work in Bayesian optimization emphasizes the importance of standardized evaluation and highlights practical challenges when applying methods to real world black-box objectives~\cite{GonzalezDuque2024NeurIPS}.

The Tree Structured Parzen Estimator is a widely used Bayesian optimization method for structured and bounded search spaces~\cite{Bergstra2011TPE,Watanabe2023cTPE}.
Recent work extends TPE to inequality constrained optimization, which aligns with tuning tasks where latency or resource limits must be satisfied~\cite{Watanabe2023cTPE}.
These developments support using TPE as a reasonable baseline for constraint-aware, budget limited tuning problems~\cite{Watanabe2023cTPE,Diessner2025Thesis}.

This paper focuses on robust and constraint-aware tuning of BBRv2 parameters for QUIC in TI style scenarios~\cite{ITUY3149, 2chaudhari2025enabling}.
The approach combines scenario robustness with hard limits on tail latency and loss and uses sample-efficient black-box optimization under expensive evaluations~\cite{ITUY3149,Watanabe2023cTPE,Diessner2025Thesis}.
The evaluation compares the tuned configuration against widely used baselines and reports metrics that directly reflect latency stability and reliability priorities~\cite{Bujari2024QUICCCA,wang2025expert}.

\section{Problem Formulation and Optimization Framework}
\label{sec:proposed_extension}

Building on prior work on utility-guided BBRv2 tuning for QUIC under tactile Internet requirements~\cite{Hanif2026BBRv2TI}, this section reformulates the tuning problem as a robust optimization task over a set of network scenarios rather than a single operating point. The objective is to identify one parameter configuration that yields stable performance across variations in delay, bandwidth, queue behavior, and loss. Since performance can only be observed by running QUIC experiments, the resulting optimization problem is treated as a noisy and expensive black box problem. To address this setting, the framework combines bounded parameter search, scenario-level aggregation, constraint handling, Bayesian optimization based on the Tree Structured Parzen Estimator, and a two-stage evaluation procedure.
\subsection{Decision Variables and Feasible Set}
\label{subsec:theta_set}

The tuning vector contains eleven BBRv2 control parameters,
following the parameterization introduced in~\cite{Hanif2026BBRv2TI}:
\begin{equation}
\boldsymbol{\theta}=
[g_c,g_p,\beta,L_{\mathrm{th}},q_r,q_d,
f_{\mathrm{bw}},L_{\mathrm{full}},h_i,\alpha,\delta]^{\mathsf T},
\qquad d=11.
\end{equation}

Table~\ref{tab:bbr_params} summarizes the parameter roles,
bounded search domains, and the final tuned configuration
\(\boldsymbol{\theta}^{\star}\) used in the evaluation.

\begin{table}[t]
\centering
\caption{BBRv2 tuning parameters, search ranges, and selected values.}
\label{tab:bbr_params}
\scriptsize
\setlength{\tabcolsep}{2.0pt}
\begin{tabular}{c l c c}
\toprule
Sym. & Control parameter & Search range & $\theta^\star$\\
\midrule
$g_c$ & Startup CWND gain          & [1.0, 3.0]    & 1.50\\
$g_p$ & Startup pacing gain        & [1.0, 3.0]    & 1.80\\
$\beta$ & Loss-recovery factor     & [0.5, 0.9]    & 0.70\\
$L_{\rm th}$ & STARTUP loss events & [1, 8]        & 3\\
$q_r$ & ProbeRTT interval (ms)     & [200, 1000]   & 500\\
$q_d$ & ProbeRTT duration (ms)     & [5, 100]      & 20\\
$f_{\rm bw}$ & Full-BW threshold   & [0.85, 1.10]  & 1.10\\
$L_{\rm full}$ & Full-BW rounds    & [1, 10]       & 3\\
$h_i$ & Inflight control factor    & [1.0, 2.5]    & 1.50\\
$\alpha$ & ProbeBW gain factor     & [0.5, 1.5]    & 1.10\\
$\delta$ & CWND smoothing factor   & [0.0, 1.0]    & 0.70\\
\bottomrule
\end{tabular}
\end{table}

Each parameter is restricted to a bounded interval to avoid
unstable or impractical configurations. The feasible search
domain is therefore
\begin{equation}
\Theta=
\left\{
\boldsymbol{\theta}\in\mathbb{R}^{11}:
\boldsymbol{\theta}^{\min}
\leq
\boldsymbol{\theta}
\leq
\boldsymbol{\theta}^{\max}
\right\},
\end{equation}
where the inequalities are applied elementwise.

\subsection{Scenario Set and Experimental Oracle}
\label{subsec:scenarios_eval}

The tuned configuration is evaluated across three network
conditions representing increasing impairment levels. Let
\(M=3\) and define the scenario set as
\begin{equation}
\mathcal{S}=\{1,2,3\}.
\end{equation}
Each scenario \(s\in\mathcal{S}\) is characterized by the
network impairment tuple
\begin{equation}
\Xi_s=(D_s,J_s^{\mathrm{net}},L_s^{\mathrm{net}}),
\end{equation}
where \(D_s\) denotes the configured network delay,
\(J_s^{\mathrm{net}}\) the delay variation, and
\(L_s^{\mathrm{net}}\) the configured packet loss rate.
Table~\ref{tab:scenarios} summarizes the experimental
profiles used throughout the evaluation.

\begin{table}[t]
\centering
\caption{Network impairment scenarios used in the evaluation.}
\label{tab:scenarios}
\scriptsize
\setlength{\tabcolsep}{5pt}
\begin{tabular}{lccc}
\toprule
Scenario & Delay & Jitter & Loss \\
\midrule
\texttt{S\_low}  & 5 ms  & 1 ms  & 0.5\% \\
\texttt{S\_mid}  & 20 ms & 5 ms  & 1.0\% \\
\texttt{S\_high} & 50 ms & 10 ms & 2.0\% \\
\bottomrule
\end{tabular}
\end{table}

All scenarios are assigned equal importance in the robust
objective; therefore,
\begin{equation}
\pi_1=\pi_2=\pi_3=\frac{1}{3},
\qquad
\sum_{s=1}^{M}\pi_s=1.
\end{equation}

For a given parameter vector \(\boldsymbol{\theta}\) and
scenario \(s\), performance is obtained by executing the
\texttt{quiche} client-server experiment under the corresponding
Linux network-emulation profile. Hence, the mapping
\((\boldsymbol{\theta},s)\) to the measured performance metrics
is treated as an experimental black-box function.
\subsection{Performance Measures}
\label{subsec:metrics_def}

For scenario \(s\) and repetition index \(r\), the experiment returns a vector of measured quantities
\begin{equation}
\boldsymbol{m}_{s,r}(\boldsymbol{\theta})
=
\begin{bmatrix}
T_{s,r}(\boldsymbol{\theta})\\
R^{95}_{s,r}(\boldsymbol{\theta})\\
L_{s,r}(\boldsymbol{\theta})\\
J_{s,r}(\boldsymbol{\theta})
\end{bmatrix},
\end{equation}
where the components denote goodput, RTT 95th percentile, packet loss rate, and jitter penalty, respectively.

\textbf{Goodput:}
Let \(U_{s,r}(\boldsymbol{\theta})\) be the number of delivered application bytes and let \(t_{s,r}(\boldsymbol{\theta})\) be the transfer duration in seconds. Goodput in Mbit/s is defined as
\begin{equation}
T_{s,r}(\boldsymbol{\theta})
=
\frac{8\,U_{s,r}(\boldsymbol{\theta})}{10^{6}\,t_{s,r}(\boldsymbol{\theta})}.
\end{equation}

\textbf{Tail latency:}
Let \(\mathcal{R}_{s,r}(\boldsymbol{\theta})\) denote the set of RTT samples collected during a run. The tail latency metric is defined by
\begin{equation}
R^{95}_{s,r}(\boldsymbol{\theta})
=
\mathrm{Percentile}_{95}
\big(
\mathcal{R}_{s,r}(\boldsymbol{\theta})
\big).
\end{equation}
This quantity is used because delay sensitive interactive systems are often more affected by high latency excursions than by average delay alone.

\textbf{Loss rate:}
Let \(N^{\mathrm{lost}}_{s,r}(\boldsymbol{\theta})\) be the number of packets declared lost, and let \(N^{\mathrm{sent}}_{s,r}(\boldsymbol{\theta})\) be the number of transmitted packets. The loss rate is
\begin{equation}
L_{s,r}(\boldsymbol{\theta})
=
\frac{N^{\mathrm{lost}}_{s,r}(\boldsymbol{\theta})}
{N^{\mathrm{sent}}_{s,r}(\boldsymbol{\theta})}.
\end{equation}

\textbf{Jitter penalty:}
To quantify short term delay instability, an estimated queueing delay is derived from RTT samples. For scenario \(s\), define
\begin{equation}
q(t)=\max\{0,\mathrm{RTT}(t)-D_s\},
\end{equation}
where \(D_s\) denotes the configured baseline network delay
for scenario \(s\). If \(\Delta t\) is the sampling interval, then the magnitude of successive queueing delay variation is
\begin{equation}
\Delta q(t)=|q(t)-q(t-\Delta t)|.
\end{equation}
The jitter penalty is then defined as the sample mean of these variations:
\begin{equation}
J_{s,r}(\boldsymbol{\theta})
=
\mathrm{Mean}\big(\Delta q(t)\big).
\end{equation}
Smaller values indicate smoother delay evolution and less queue instability.

\subsection{Repeated Evaluation and Scenario Aggregation}
\label{subsec:averaging_noise}

Because transport measurements vary across runs due to timing effects, protocol state evolution, and stochastic network behavior, a single run does not provide a sufficiently reliable estimate of performance. Accordingly, each scenario is repeated multiple times. Let \(N_s\) denote the number of repetitions associated with scenario \(s\).

The scenario-level average performance vector is defined as
\begin{equation}
\bar{\boldsymbol{m}}_s(\boldsymbol{\theta})
=
\frac{1}{N_s}
\sum_{r=1}^{N_s}
\boldsymbol{m}_{s,r}(\boldsymbol{\theta})
=
\begin{bmatrix}
\bar{T}_s(\boldsymbol{\theta})\\
\bar{R}^{95}_s(\boldsymbol{\theta})\\
\bar{L}_s(\boldsymbol{\theta})\\
\bar{J}_s(\boldsymbol{\theta})
\end{bmatrix}.
\end{equation}

These averaged quantities are used in the optimization stage in order to reduce the effect of run-to-run variability and provide more stable estimates of scenario-level behavior.

\subsection{Robust Constrained Optimization Problem}
\label{subsec:constraints_obj}

The tuning objective seeks a single BBRv2 configuration that
maintains high goodput while satisfying tail-latency and loss
requirements across all network scenarios. For each
\(s\in\mathcal{S}\), feasibility requires
\(\bar{R}^{95}_{s}(\boldsymbol{\theta})\le R_{\mathrm{limit}}\)
and
\(\bar{L}_{s}(\boldsymbol{\theta})\le L_{\mathrm{limit}}\).

Using the scenario weights defined in
Section~\ref{subsec:scenarios_eval}, the robust optimization
problem is formulated as
\begin{equation}
\begin{aligned}
\max_{\boldsymbol{\theta}\in\Theta}\quad
& \sum_{s=1}^{M}\pi_s
  \bar{T}_{s}(\boldsymbol{\theta}) \\
\mathrm{subject\ to}\quad
& \bar{R}^{95}_{s}(\boldsymbol{\theta})
  \le R_{\mathrm{limit}},
  && \forall s\in\mathcal{S},\\
& \bar{L}_{s}(\boldsymbol{\theta})
  \le L_{\mathrm{limit}},
  && \forall s\in\mathcal{S},\\
& \pi_s\ge 0,
  && \forall s\in\mathcal{S},\\
& \sum_{s=1}^{M}\pi_s=1.
\end{aligned}
\label{eq:robust_problem}
\end{equation}

For the three scenarios considered in this study,
\(\pi_1=\pi_2=\pi_3=1/3\), so each impairment level contributes
equally to the optimization objective.
\subsection{Penalty-Based Black-Box Reformulation}
\label{subsec:blackbox_form}

Since the objective and constraints are available only through
experimental evaluations, the constrained problem is converted
into a scalar black-box objective. To avoid directly combining
quantities with different physical units, the RTT and loss
violations are expressed relative to their respective limits.

For scenario \(s\), define the normalized constraint violations as
\begin{equation}
v^{R}_{s}(\boldsymbol{\theta})
=
\max\left\{
0,
\frac{\bar{R}^{95}_{s}(\boldsymbol{\theta})
      -R_{\mathrm{limit}}}
     {R_{\mathrm{limit}}}
\right\},
\end{equation}

\begin{equation}
v^{L}_{s}(\boldsymbol{\theta})
=
\max\left\{
0,
\frac{\bar{L}_{s}(\boldsymbol{\theta})
      -L_{\mathrm{limit}}}
     {L_{\mathrm{limit}}}
\right\}.
\end{equation}

The aggregate constraint violation is then
\begin{equation}
V(\boldsymbol{\theta})
=
\sum_{s=1}^{M}\pi_s
\left[
v^{R}_{s}(\boldsymbol{\theta})
+
v^{L}_{s}(\boldsymbol{\theta})
\right],
\end{equation}
which is dimensionless and assigns comparable influence to
delay and loss violations.

The scalar objective minimized by the optimizer is
\begin{equation}
J(\boldsymbol{\theta})
=
-
\sum_{s=1}^{M}\pi_s
\bar{T}_{s}(\boldsymbol{\theta})
+
\lambda V(\boldsymbol{\theta}),
\label{eq:penalty_score}
\end{equation}
where \(\lambda>0\) controls the cost of violating the
Tactile Internet constraints. Smaller values of
\(J(\boldsymbol{\theta})\) correspond to configurations with
higher goodput and fewer constraint violations.

\subsection{Bayesian Search via Tree-Structured Parzen Estimation}
\label{subsec:why_tpe}

The objective in~\eqref{eq:penalty_score} is expensive and noisy
because each evaluation requires repeated QUIC experiments over
the scenario set. Therefore, a sample-efficient sequential search
strategy is used. The Tree-Structured Parzen Estimator (TPE)
constructs probabilistic models from previously evaluated
configurations rather than requiring gradients.

After an initial set of evaluations, the observed configurations
are separated according to an objective-value threshold into
promising and non-promising sets. TPE models these regions using
the conditional densities
\begin{equation}
\ell(\boldsymbol{\theta})
=
p(\boldsymbol{\theta}\mid J(\boldsymbol{\theta})\le J_{\gamma}),
\qquad
g(\boldsymbol{\theta})
=
p(\boldsymbol{\theta}\mid J(\boldsymbol{\theta})>J_{\gamma}),
\end{equation}
where \(J_{\gamma}\) denotes the selected objective quantile.
Candidate parameter vectors are sampled from
\(\ell(\boldsymbol{\theta})\) and ranked according to the density
ratio
\begin{equation}
\boldsymbol{\theta}_{i}
=
\arg\max_{\boldsymbol{\theta}}
\frac{\ell(\boldsymbol{\theta})}
     {g(\boldsymbol{\theta})}.
\end{equation}
The selected candidate is evaluated experimentally, its objective
value is added to the observation history, and the density models
are updated before the next iteration. This sequential procedure
concentrates evaluations in promising regions while retaining
exploration of the bounded parameter space.

\subsection{Two-Stage Evaluation Protocol}
\label{subsec:two_stage}

To limit the cost of repeated network experiments, candidate
configurations are evaluated using a two-stage protocol. The
implementation uses \(R_{\mathrm{limit}}=125~\mathrm{ms}\),
\(L_{\mathrm{limit}}=0.025\), and penalty coefficient
\(\lambda=25~\mathrm{Mbit/s}\). The Stage-A search budget is \(B=150\)
candidate configurations, and the best \(K=5\) candidates are
retained for confirmation.

Stage A evaluates each candidate once under every scenario,
providing a low-cost estimate for guiding the TPE search.
Stage B reevaluates each of the \(K\) selected candidates using
five repetitions per scenario, and the resulting averaged
measurements are used for final ranking. This procedure reduces
the cost of exploring the parameter space while applying higher
evaluation fidelity to the most promising configurations.

\subsection{Optimization Procedure}
\label{subsec:procedure}

Algorithm~\ref{alg:final_method} summarizes the complete tuning workflow.

\begin{algorithm}[t]
\caption{Robust constraint-aware tuning using TPE with a two-stage evaluation protocol}
\label{alg:final_method}
\begin{algorithmic}[1]

\STATE \textbf{Input:} feasible set \(\Theta\), scenario set \(\mathcal{S}\), scenario weights \(\boldsymbol{\pi}\)
\STATE \textbf{Input:} limits \(R_{\mathrm{limit}}\), \(L_{\mathrm{limit}}\), penalty coefficient \(\lambda\)
\STATE \textbf{Input:} Stage-A budget \(B\), Stage-B candidate count \(K\)
\STATE Initialize observation history \(\mathcal{D}\leftarrow\emptyset\)

\FOR{\(i=1\) to \(B\)}
    \STATE Propose \(\boldsymbol{\theta}_{i}\) using the TPE density ratio
    \STATE Evaluate \(\boldsymbol{\theta}_{i}\) under Stage A
    \STATE Compute \(J(\boldsymbol{\theta}_{i})\) using \eqref{eq:penalty_score}
    \STATE Store \((\boldsymbol{\theta}_{i},J(\boldsymbol{\theta}_{i}))\) in \(\mathcal{D}\)
\ENDFOR

\STATE Select the \(K\) candidates with the smallest Stage-A objective values

\FOR{each selected candidate \(\boldsymbol{\theta}\)}
    \STATE Reevaluate \(\boldsymbol{\theta}\) under Stage B
    \STATE Recompute \(J(\boldsymbol{\theta})\)
\ENDFOR

\STATE \textbf{Output:} \(\boldsymbol{\theta}^{\star}\) with the smallest Stage-B objective value

\end{algorithmic}
\end{algorithm}

\section{Results}
\label{sec:results}

Experiments were performed using \texttt{quiche} under Linux network emulation. Three scenarios, \texttt{S\_low}, \texttt{S\_mid}, and \texttt{S\_high}, represent increasing impairment levels. The evaluated algorithms are \texttt{reno}, \texttt{cubic}, \texttt{bbr}, \texttt{bbr2}, and the tuned method \texttt{bbrTIOB}. Results are reported as mean and standard deviation over repeated runs using goodput, RTT 95th percentile, loss rate, jitter penalty, a combined score, and the RTT CDF.

\subsection{Performance Across Network Scenarios}
\label{subsec:results_main}

\begin{figure}[!tb]
\centering
\includegraphics[width=0.92\linewidth]{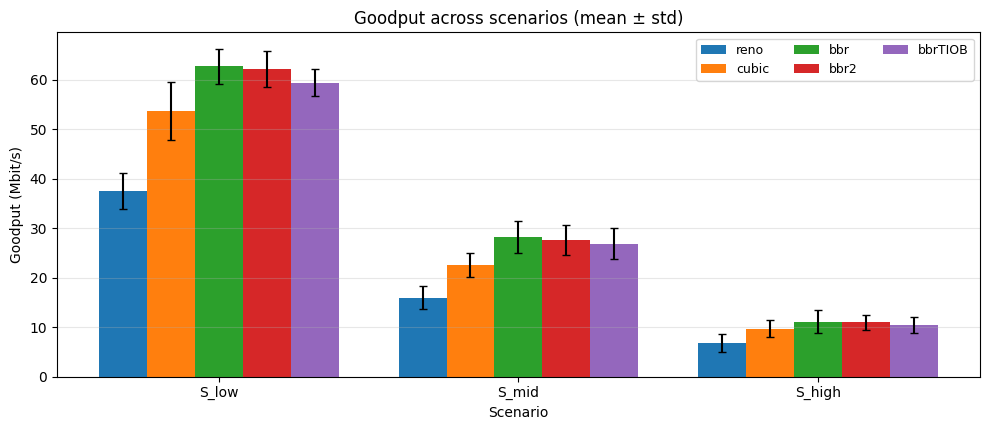}
\caption{Goodput across scenarios shown as mean and standard deviation.}
\label{fig:goodput_mean_std}
\end{figure}

\paragraph{Goodput}
Figure~\ref{fig:goodput_mean_std} shows that goodput decreases as impairment increases. Among the baselines, \texttt{bbr} and \texttt{bbr2} achieve higher goodput than \texttt{reno} and \texttt{cubic}. The tuned method \texttt{bbrTIOB} remains close to \texttt{bbr} and \texttt{bbr2} across all scenarios, indicating that the gains in latency and stability are not obtained at the cost of a major throughput penalty.

\begin{figure}[!tb]
\centering
\includegraphics[width=0.92\linewidth]{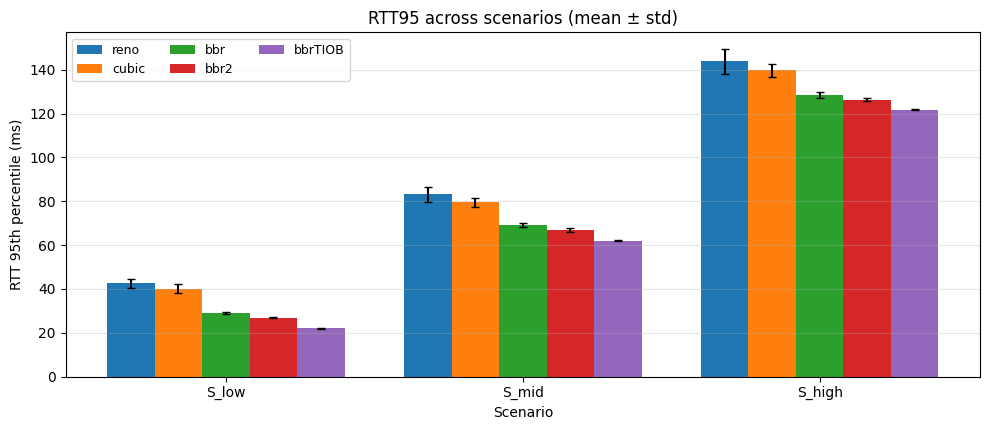}
\caption{RTT 95th percentile across scenarios shown as mean and standard deviation.}
\label{fig:rtt95_mean_std}
\end{figure}

\paragraph{Tail latency}
Figure~\ref{fig:rtt95_mean_std} reports RTT 95th percentile. Tail latency increases with impairment, but \texttt{bbrTIOB} consistently achieves the lowest values across all scenarios. This indicates improved control of queue buildup and delay spikes relative to the baseline methods.

\begin{figure}[!tb]
\centering
\includegraphics[width=0.92\linewidth]{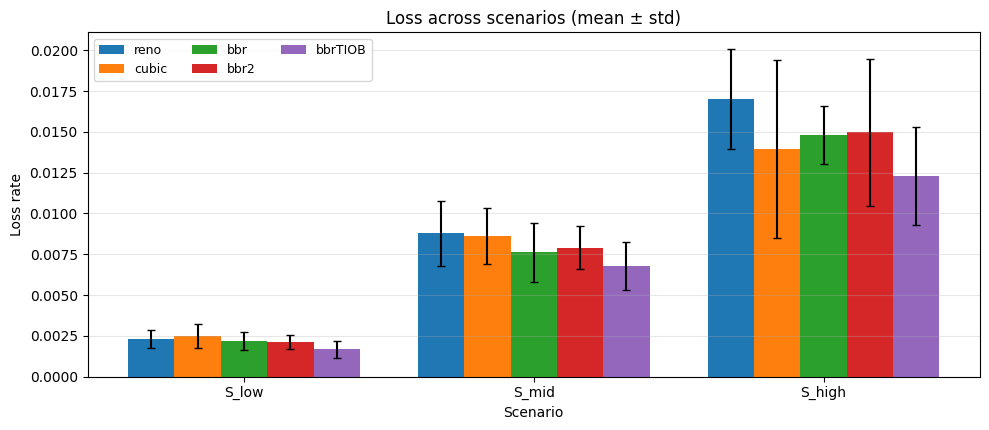}
\caption{Loss rate across scenarios shown as mean and standard deviation.}
\label{fig:loss_mean_std}
\end{figure}

\paragraph{Loss}
Figure~\ref{fig:loss_mean_std} shows that loss increases from \texttt{S\_low} to \texttt{S\_high}. The tuned method yields lower loss than the baselines, with the clearest gain in the high-impairment case, suggesting more stable and less aggressive sending behavior.

\begin{figure}[!tb]
\centering
\includegraphics[width=0.92\linewidth]{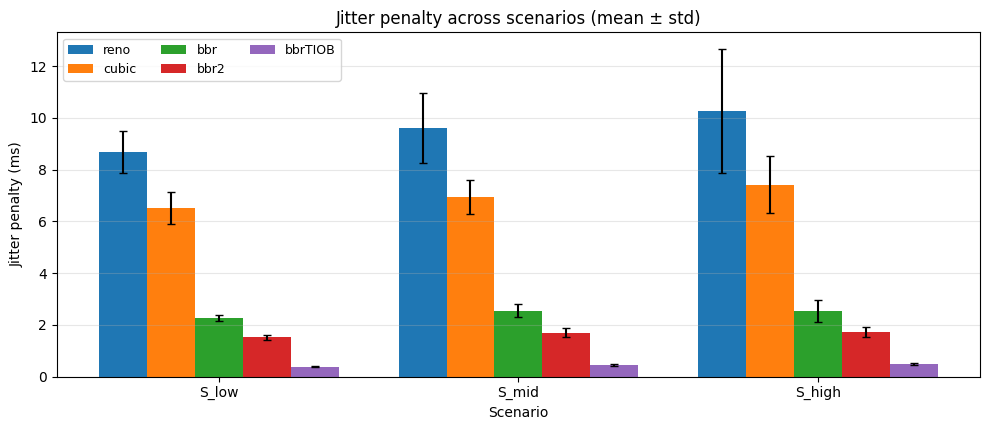}
\caption{Jitter penalty across scenarios shown as mean and standard deviation.}
\label{fig:jitter_mean_std}
\end{figure}

\paragraph{Jitter}
Figure~\ref{fig:jitter_mean_std} reports the jitter penalty, where lower values indicate smoother delay behavior. \texttt{bbrTIOB} achieves the lowest jitter penalty in all scenarios, showing improved delay stability.

\begin{figure}[!tb]
\centering
\includegraphics[width=0.92\linewidth]{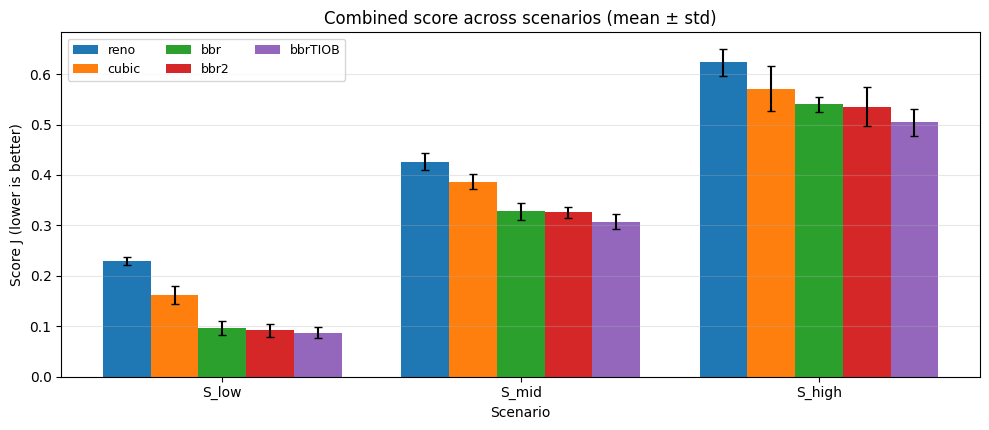}
\caption{Combined score across scenarios shown as mean and standard deviation. Smaller values indicate better overall performance.}
\label{fig:score_mean_std}
\end{figure}

\paragraph{Combined score}
Figure~\ref{fig:score_mean_std} summarizes the tradeoff among goodput, RTT 95th percentile, loss, and jitter. \texttt{bbrTIOB} achieves the lowest score in the set of scenarios, indicating the best overall balance among the metrics evaluated.

\subsection{Distributional and Optimization Analysis}
\label{subsec:results_additional}

\begin{figure}[!t]
\centering
\includegraphics[width=0.88\linewidth]{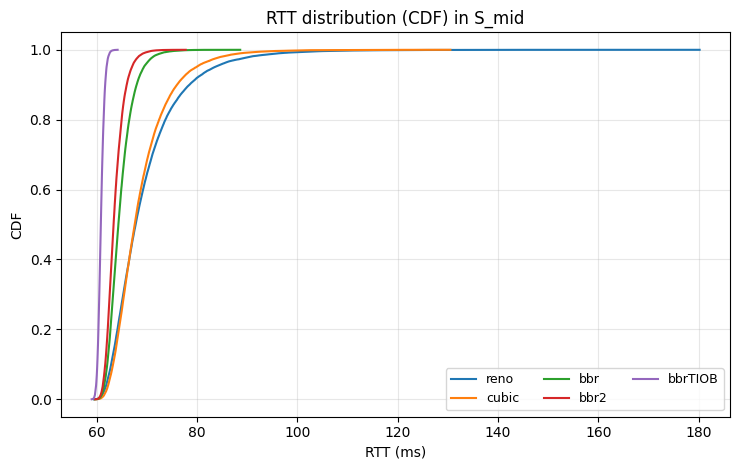}
\caption{RTT CDF in \texttt{S\_mid}. A curve farther to the left indicates lower RTT values.}
\label{fig:rtt_cdf_smid}
\end{figure}

\paragraph{RTT CDF}
Figure~\ref{fig:rtt_cdf_smid} shows the RTT distribution in \texttt{S\_mid}. The \texttt{bbrTIOB} curve is shifted left relative to the baselines, confirming lower RTT over a large fraction of samples and supporting the tail latency results.

\begin{figure}[!t]
\centering
\includegraphics[width=0.88\linewidth]{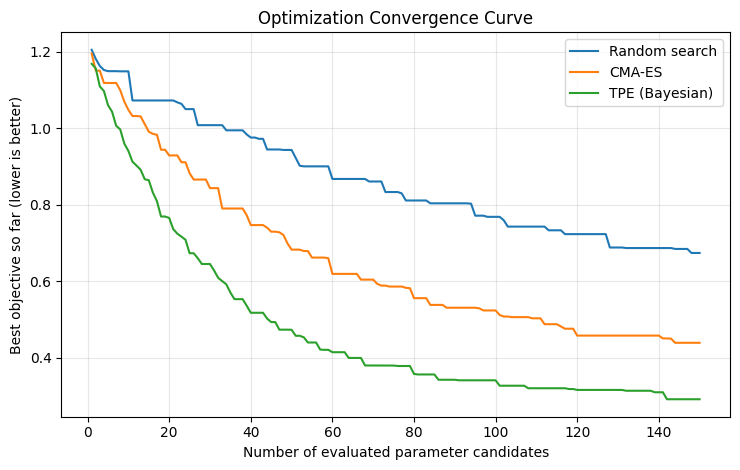}
\caption{Optimization convergence. Lower objective values indicate better candidate configurations.}
\label{fig:convergence}
\end{figure}

\paragraph{Convergence}
Figure~\ref{fig:convergence} compares TPE, CMA-ES, and random
search under the same candidate-evaluation budget. TPE reaches
lower objective values more rapidly and maintains the lowest
best-so-far value toward the end of the search, indicating more
sample-efficient exploration of the tuning space. Across \texttt{S\_low}, \texttt{S\_mid}, and \texttt{S\_high}, \texttt{bbrTIOB} improves tail latency, jitter, and loss while maintaining competitive goodput, further supporting the proposed tuning framework.
The optimizer comparison is therefore reported through search
convergence, while the protocol-level evaluation uses the final
configuration selected by the proposed TPE framework.
\section{Conclusion}
\label{sec:conclusion}
This paper presented a robust and constraint-aware tuning framework for BBRv2 in QUIC for Tactile Internet-style communication. The problem was formulated as a black-box optimization task across multiple network scenarios, and Bayesian optimization with the Tree Structured Parzen Estimator was used to search the parameter space efficiently under noisy experimental evaluations. Results show that the tuned configuration improves tail latency, jitter, and loss while maintaining competitive goodput compared with standard baselines. These findings suggest that careful tuning can make BBRv2 more suitable for delay-sensitive QUIC applications. The present evaluation focuses on controlled Linux network
emulation; future work will incorporate radio, device-processing, and complete end-to-end tactile-loop effects in a physical testbed.

\section*{Acknowledgment}
The work in this paper was partially supported by the
Palmer Department Chair Endowment.

\bibliographystyle{IEEEtran}
\bibliography{references}

\end{document}